\documentclass[fleqn,twoside]{article}
\usepackage{espcrc2}

\usepackage{graphicx}
\usepackage[figuresright]{rotating}

\newcommand{\AmS}{{\protect\the\textfont2
  A\kern-.1667em\lower.5ex\hbox{M}\kern-.125emS}}

\title{Phenomenology of static-light mesons from unquenched lattice
       QCD calculations.}

\author{UKQCD Collaboration, Craig~McNeile\address[LIV]{Department of Mathematical Sciences, 
University of\ Liverpool, L69 3BX, UK}
Chris~Michael\addressmark[LIV], and
Gavin Thompson\addressmark[LIV]
}

\begin{document}

\begin{abstract}
We present results for the static-light meson from 
unquenched lattice QCD. The unquenched gauge configurations
were generated using the non-perturbatively improved clover
action. At the fixed lattice spacing of 0.1 fm the lightest sea quark
mass used is a third of the strange quark mass. A comparison is made
between heavy-light chiral perturbation theory and the
$f_B^{\textit{static}}$
decay constant. The mass of the bottom quark is also reported:
$\overline{m_b} (\overline{m_b} )$ =
4.25(2)(11) GeV, where the first error is statistical and 
the last error is the systematic uncertainty.

\vspace{1pc}
\end{abstract}

\maketitle

\section{INTRODUCTION}

There are a number of important issues in heavy quark phenomenology
that can be addressed using the static quark formulation for the heavy
quark.  The mass of the bottom quark is a fundamental parameter of the
standard model. As noted by
Gimenez et al~\cite{Gimenez:2000cj}, the error on the mass of the
bottom quark due to the use of static quarks for the $b$ quark is only
of order 30 MeV~\cite{Gimenez:2000cj}.

The heavy flavour experiments such as BaBar, BELLE, CDF, and D0 are
extensively testing the CKM matrix formalism.  There are a number of
non-perturbative QCD matrix elements that must also be computed to
test the CKM formalism. In particular, the ratio of the decay
constants of the $B_s$ to $B$ mesons ($\frac{f_{B_s}}{f_B}$) is a
crucial QCD quantity for the unitarity checks of the CKM matrix.

\section{COMPUTATIONAL DETAILS}

The basis of our calculation is unquenched 
gauge configurations generated with 
the non-perturbatively improved clover
action and the Wilson gauge action.
The lattice parameters are: volume $16^3 \; 32$, 
$\beta$ = 5.2, and  the 
clover coefficient was the non-perturbative value of 2.0171.
The full details of the action and results
on the hadron spectroscopy have 
been published~\cite{Allton:2001sk}.

UKQCD has already published~\cite{Green:2003zz}
an extensive analysis
of the spectrum of static-light mesons.

The $f_B$ decay constant is extracted from 
the amplitudes in the two point correlator.
\begin{eqnarray}
C(t) & = &  \sum_{x} 
\langle 0 \mid 
A_4(x,t)
\Phi_B^\dagger(x,0) 
\mid 0 \rangle \\
     & \rightarrow & Z_L Z_{\Phi_B} \exp( - a{ \cal E } t)
\end{eqnarray}
where 
$\Phi_B$ is the interpolating operator for  static-light mesons.
We used all-to-all propagators and fuzzed sources to get accurate
correlators~\cite{Green:2003zz}. We fit a 3 exponential model to a 5
by 5 smearing matrix.

The $f_B$ decay constant is defined by the matrix element 
below:
\begin{equation}
\langle 0 \mid A_\mu \mid B(p) \rangle = i p_\mu f_B
\end{equation}
The axial current is improved using the 
ALPHA formulation~\cite{Kurth:2000ki},
including the correction term introduced by 
Morningstar and 
Shigemitsu~\cite{Morningstar:1998yx}.
The connection between the decay constant and the 
amplitude is.
\begin{equation}
f_B^{static} = Z_L \sqrt{ \frac{2}{M_B} } Z_A^{static}
\end{equation}
where $Z_A^{static}$ is the renormalisation factor.
Here we explore the sea quark dependence of $Z_L$.
We  have checked that the variation of
$Z_A^{static}$ with sea quark mass is small.

\section{THE MASS OF THE BOTTOM QUARK}

Full details of 
our calculation of the bottom quark mass
have been reported 
in~\cite{McNeile:2004cb}.

The parameters $\kappa_{sea} = 0.1355$
$\kappa_{val} = 0.1350$ were used for the central
value.
The quantity ${\cal E}$, from the lattice
calculation, contains an unphysical  $\frac{1}{a}$ divergence ($\delta
m$) that must be subtracted off to obtain the physical binding 
energy ($\Lambda_{static}$).
\begin{equation}
\Lambda_{static} = {\cal E } - \delta m
\end{equation}
 The pole quark mass is determined from
\begin{equation}
m_b^{pole} = M_{B_s} - \Lambda_{static}
\end{equation}
The physical value~\cite{Hagiwara:2002fs} of the meson mass
$M_{B_s}$  (5.369 GeV) is used.

In the static theory  $\delta m$ has been
calculated to two loops by 
Martinelli and Sachrajda~\cite{Martinelli:1998vt}.
The pole mass is
converted to $\overline{MS}$ using continuum perturbation 
theory~\cite{Gray:1990yh}.
\begin{equation}
m_b^{\overline{MS}}(\mu) = 
Z_{pm}(\mu)
m_b^{pole}
 + O(1/m_b)
\label{eq:masterEq}
\end{equation}
The lattice matching is only done to $O(\alpha^2)$, hence we convert
the pole mass to $\overline{MS}$ at the same order, using a consistent
coupling, so the differences in the series are physical. This avoids
problems with renormalons.

Our final result for $\overline{m_b} (\overline{m_b})$ 
(in GeV) is
\begin{equation}
4.25  \pm 0.02  \pm 0.03  \pm 0.03 \pm 0.08 \pm 0.06 
\end{equation}
where the errors are (from left to right):
statistical, perturbative, neglect
of $1/m_b$ terms, ambiguities in the choice of
lattice spacing, and error in the choice of the mass
of the strange quark.

Our result is consistent with that from
Gimenez et al.~\cite{Gimenez:2000cj} 
$(\bar{m}_{b}(\bar{m}_{b}) = 4.26 \pm 0.09 \; {\rm GeV})$
from an unquenched QCD calculation.

\section{CHIRAL LOGS IN THE HEAVY-LIGHT DECAY CONSTANT}

The error on the ratio of the $\frac{f_{B_s}}{f_B}$ has recently been
increased, because the chiral log term has not been observed
in lattice data~\cite{Kronfeld:2002ab}. For example, 
the JLQCD~\cite{Aoki:2003xb} 
collaboration quote 
$f_{B_s}/f_{B_d}=1.13(3)(^{+13}_{-\ 2})$,
where the first error is statistical and the second
error is from the systematic uncertainties.
The dominant systematic uncertainty in JLQCD's result
is from the chiral extrapolation.
All lattice calculations, apart from some
preliminary evidence from one
~\cite{Wingate:2003gm},
have only seen linear 
dependence of the heavy-light decay constant 
on the quark mass.

The one loop correction, 
in heavy-light chiral perturbation theory,
to the static-light decay
constant is~\cite{Aoki:2003xb}
\begin{equation}
\frac
{\Phi_{f_{B_d}}}
{\Phi_{f_{B_d}^0}}
= 1 - \frac{3(1 + 3 g^2) }{4} \frac{ m_\pi^2 } { (4 \pi f )^2 }
\log( \frac{ m_\pi^2  } { \mu^2} )
\label{eq:heavyLight}
\end{equation}
where $g$ is  the $B^{*}B\pi$ coupling
and $\Phi_{f_{B_d}} \equiv f_{B_d} \sqrt{M_{B_d}}$.
The coupling $g$ has recently been measured at CLEO.
It has also been determined from quenched
lattice QCD.

\begin{figure}[t]
\begin{center}
\leavevmode
\includegraphics[scale=0.3,clip,angle=270]{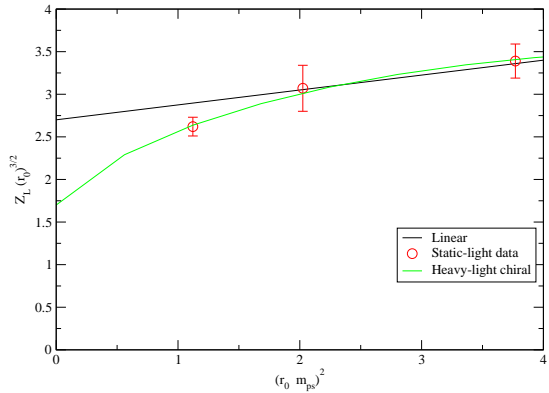}
\end{center}
\caption[]{\label{eq:heavyLog} {
Static-light decay constant as a function of quark mass.
}}
\end{figure}

In figure~\ref{eq:heavyLog} 
$Z_L$ using $\kappa_{sea}$ = 0.1350, 0.1355, and 0.1358
is plotted. At $\kappa_{sea}$ = 0.1358, 138 configurations
were used. The other two $\kappa$ values used
ensemble sizes of 60.
JLQCD~\cite{Aoki:2003xb} 
reported that the heavy-light decay constant had only linear
dependence on quark mass.
The new data at $\kappa_{sea}$ = 0.1358
does show some deviation from linearity. Although the 
figure~\ref{eq:heavyLog} is encouraging -- the deviation is 
not statistically significant. The curve is 
the 
expression from
chiral perturbation theory (equation~\ref{eq:heavyLight})
drawn for illustration, rather the
result of a fit. 

As has been noted by many 
authors~\cite{Bernard:2002pc,Becirevic:2002mh}
the chiral log structure 
of $f_\pi$  and $f_B$ are rather similar. 
Hence UKQCD's claim to see the effect of the 
chiral logs in $f_\pi$~\cite{Allton:2004qq}
suggests that there should be deviations
from linear dependence of $f_{B_s}^{static}$
on the quark mass.

The value of the lightest pion in 
our calculation is 
roughly 420 MeV~\cite{Allton:2004qq}.
The different treatments of the heavy-light chiral
perturbation theory
of Sanz-Cillero et al.~\cite{Sanz-Cillero:2003fq}
show that a deviation of linearity 
in quark mass for $f_B$
is expected at these
pion masses.

In UKQCD's work on chiral logs in the pion decay constant,
there was a concern about finite volume 
effects~\cite{Allton:2004qq}.
At $\kappa_{sea}$ = 0.1358
it was argued from chiral
perturbation theory that the finite volume effects
were of the order of 8\% in $f_\pi$. 
A similar order of magnitude effect
was also estimated by Colangelo and Haefeli~\cite{Colangelo:2004xr}.
The volume of the lightest data set is $(1.5 \; \mbox{fm})^3$.
Recently Arndt and Lin~\cite{Arndt:2004bg} have studied the 
effect of the finite volume on the ratio of heavy light decay 
constants and bag parameters. For a pion mass around 400 MeV
on box size of $(1.6 \; \mbox{fm})^3$,
the finite volume effect in the ratio $\frac{f_{B_s}}{f_B}$
is 0.006. This suggests that the finite size effects 
are small. However, the next to leading order estimate 
of finite 
size effects in $f_\pi$ was significant~\cite{Colangelo:2004xr}. 
Unfortunately, 
Colangelo and Haefeli~\cite{Colangelo:2004xr} claim that
there is not enough information to do a similar estimate for 
$f_B$.

As the aim of this work is to look for chiral logs in 
the $f_B$ decay constant that are a small effect, it is
important to reduce the statistical errors. We are already using
 all-to-all techniques and fuzzing. The number of available
gauge configurations is fixed. 
The ALPHA collaboration~\cite{DellaMorte:2003mn}
have developed 
a new variant of the static formalism that reduces the 
$1/a$ mass renormalisation that is thought to be the reason for 
the poor signal to noise ratio in static-light calculations.
We are currently running with this formalism and we expect to
report results in the future.



\end{document}